\documentclass[a4paper,fleqn]{cas-sc}

\newcommand{\printorcid}{}

\usepackage[numbers]{natbib}
\usepackage{subcaption}
\def\tsc#1{\csdef{#1}{\textsc{\lowercase{#1}}\xspace}}
\tsc{WGM}
\tsc{QE}

\begin{document}
\let\WriteBookmarks\relax
\def\floatpagepagefraction{1}
\def\textpagefraction{.001}

\shorttitle{}    

\shortauthors{}  

\title [mode = title]{A Hierarchical Reinforcement Learning Framework for Multi-UAV Combat Using Leader-Follower Strategy}  

\author[1]{Jinhui Pang}
\ead{pangjinhui@bit.edu.cn}
\credit{Resources, Validation, Data curation}
\affiliation[1]{organization={School of computer},
            addressline={Beijing Institute of Technology}, 
            city={Beijing},
            postcode={100081}, 
            country={China}}

\author[1]{Jinglin He}
\ead{3220231372@bit.edu.cn}
\credit{Conceptualization, Methodology, Software, Writing - original draft}

\author[1]{Noureldin Mohamed Abdelaal Ahmed Mohamed}
\ead{noureldinmoamedab@gmail.com}
\credit{Software, Visualization, Writing - review \& editing}

\author[1]{Changqing Lin}
\ead{lcq2000@bit.edu.cn}
\credit{Investigation, Methodology, Writing - original draft}

\author[1]{Zhihui Zhang}
\ead{3220231441@bit.edu.cn}
\credit{Formal analysis, Data curation}

\author[2]{Xiaoshuai Hao}
\cormark[1]
\ead{xshao@baai.ac.cn}
\credit{Supervision, Writing - review \& editing}
\affiliation[2]{organization={Beijing Academy of Artificial Intelligence},
            city={Beijing},
            postcode={100862}, 
            country={China}}

\cortext[1]{Corresponding author}

\begin{abstract}
Multi-UAV air combat is a complex 
task involving multiple autonomous UAVs, an evolving field in both aerospace and artificial intelligence. 
This paper aims to enhance adversarial performance through collaborative strategies.
Previous approaches predominantly discretize the action space into 
predefined actions, limiting UAV maneuverability and complex strategy implementation.
Others simplify the problem to 1v1 combat, neglecting the cooperative dynamics among  multiple UAVs.
To address the high-dimensional challenges inherent in six-degree-of-freedom space and improve cooperation, we propose a hierarchical framework utilizing the Leader-Follower Multi-Agent Proximal Policy Optimization (LFMAPPO) strategy. 
Specifically, the framework is structured into three levels. 
The top level conducts a macro-level assessment  of the environment and guides execution policy.
The middle level determines the angle of the desired action. The bottom level generates precise action commands 
for the high-dimensional action space.
Moreover, we optimize the  state-value functions by assigning distinct roles 
with the leader-follower strategy to train the top-level policy, followers estimate the leader’s utility, promoting effective cooperation among agents. 
Additionally, the incorporation of a target selector, aligned with the UAVs' posture, assesses the threat level of targets.
Finally, simulation experiments validate the effectiveness of our proposed method.
\end{abstract}


\maketitle

\section{Introduction}
Multi-UAV air combat is a complex military operation involving multiple UAVs engaging dynamic aerial targets. Each UAV as an independent agent, collaborating with others to defeat enemy forces. Unlike single-UAV missions, multi-UAV combat achieves a "1+1>2" effect through cooperation, which includes coordinated flight maneuvers, joint strikes, and the deployment of decoy tactics~\cite{zhou2022swarm}.
Given their cost-effectiveness and suitability for high-risk missions on modern battlefields, air combat systems are progressively transitioning from human-centered control to human-assisted operations, with UAVs autonomously sensing the external environment and making independent flight decisions.

\begin{figure}[]
    \centering
    \begin{subfigure}[t]{0.3\linewidth}
        \centering
        \includegraphics[width=0.8\linewidth]{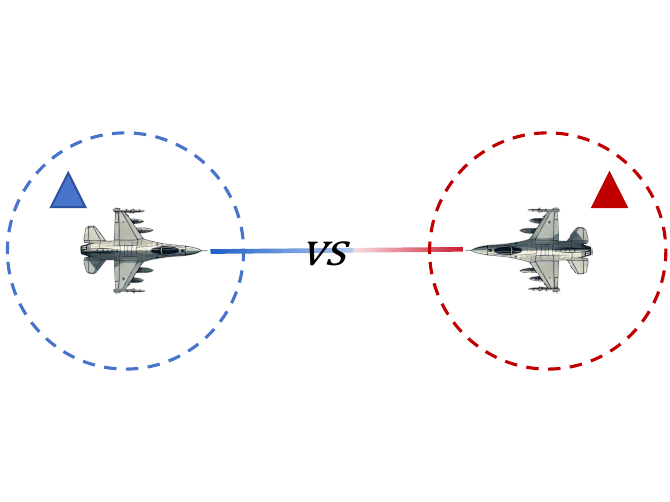}
        \caption{Single-UAV combat}
        \label{fig:image1}
    \end{subfigure}
    \hfill
    \begin{subfigure}[t]{0.3\linewidth}
        \centering
        \includegraphics[width=0.8\linewidth]{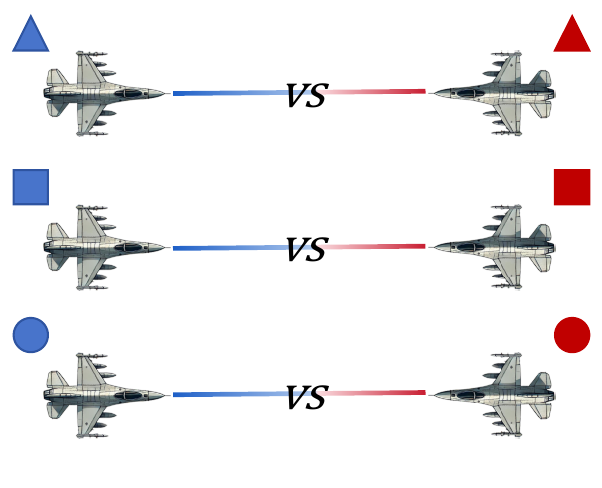}
        \caption{Multi-UAV in one-on-one combat}
        \label{fig:image2}
    \end{subfigure}
    \hfill    
    \begin{subfigure}[t]{0.3\linewidth}
        \centering
        \includegraphics[width=\linewidth]{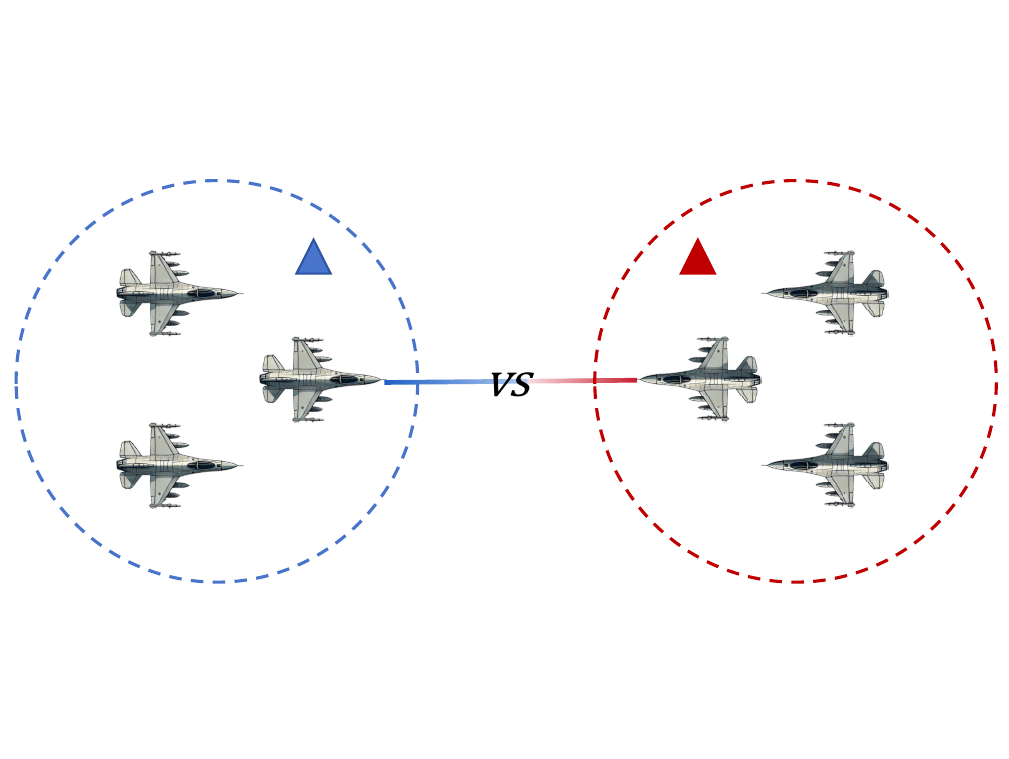}
        \caption{Cooperate multi-UAV combat}
        \label{fig:image3}
    \end{subfigure}
    
    \caption{\textbf{Three different scenarios of UAV combat.} (a) Single-UAV combat: An individual UAV engages in combat against a single opponent.
(b) Multi-UAV in one-on-one combat: Each UAV is paired with a corresponding adversary, emphasizing independent engagements without coordination between UAVs.
(c) Cooperative multi-UAV combat: A coordinated engagement where multiple UAVs work together as a team against a group of adversaries.}
    \label{fig:all-images}
\end{figure}

Rule-based methods~\cite{belkin1987cooperative,arar2013flexible,hu2021application} have guided UAV behavior in combat scenarios through predefined rules and strategies. Furthermore, some studies~\cite{burgin1988rule,jiandong2021uav,piao2023complex} have integrated expert systems to analyze pilot decision-making experiences, extracting these decisions into rule sets and compiling them into decision-making databases. However, rule-based methods exhibit limitations when faced with complex and dynamic combat environments.
Game theory~\cite{marden2018game} has been increasingly introduced into UAV decision modeling. It effectively describes interactions between agents and provides mathematical modeling tools for UAV missions and design constraints~\cite{cao2023autonomous, cruz2001game, xu2018application,ren2023cooperative}. However, game theory models face challenges in practical applications, including high modeling complexity and strong assumptions, which lead to increased computational complexity and potential deviations from actual results. 
Niu et al.~\cite{niu2023three} develope a perception-inspired learning framework to address multi-constrained collaborative planning problems.
In the intelligent air combat decision-making,  Pope et al.\cite{pope2021hierarchical}, in the DARPA AlphaDogfight close combat simulation, trained UAVs using a hierarchical structure and maximum entropy reinforcement learning, defeating an active-duty F-16 pilot. This achievement demonstrates the feasibility and effectiveness of reinforcement learning in addressing intelligent air combat decision-making problems\cite{zhang2022multi, wang2024evolutionary, wang2024deep}. 
The core of reinforcement learning lies in learning optimal decisions through continuous interaction with the environment, employing a "trial-and-error" process where agents aim to maximize cumulative rewards~\cite{dogru2024reinforcement}. This is based on the theoretical foundation of the Markov Decision Process (MDP), which effectively models the interaction behaviors of agents in air combat, making agents learn optimal strategies through environmental feedback. Different researchs have adopted various methods to address the challenges of high-dimensional complexity and task decomposition. Qiu et al.\cite{qiu2020one} simplified the three-dimensional air combat environment into a two-dimensional one to solve the air combat decision-making problem. While this approach reduced computational complexity, it also resulted in  the loss of many key features of the three-dimensional environment. In contrast, Yang et al.\cite{yang2019uav} modeled the three-dimensional air combat environment and used the Deep Deterministic Policy Gradient (DDPG) algorithm framework to solve the maneuver decision-making problem in this environment. Meanwhile, Fan et al.\cite{fan2022air} and Yang et al.\cite{yang2019maneuver} simulated UAV maneuver strategies for various air combat environments, demonstrating the flexibility of reinforcement learning methods in handling diverse task requirements.
While research on 1v1 combat scenarios is relatively mature, many multi-UAV studies often overlook UAV cooperation, instead divideding multi-UAV tasks into multiple independent 1v1 engagements, as shown in Figure~\ref{fig:all-images}.
To further optimize multi-UAV collaborative combat capabilities, Zhang et al.\cite{jiandong2021uav} implemented communication between UAVs using bidirectional recurrent neural networks, providing technical support for multi-agent cooperation. Ma et al.\cite{ma2018air} trained agents using the Deep Q-Network (DQN) algorithm and successfully defeated opponents by combining Monte Carlo Tree Search and Minimax algorithms. Hu et al.~\cite{hu2021application} introduced a situational awareness layer based on the DQN algorithm and validated its effectiveness in various air combat scenarios, further enhancing the system's perception and decision-making capabilities.

\begin{figure}[]
\centering
\includegraphics[width=0.45\columnwidth]{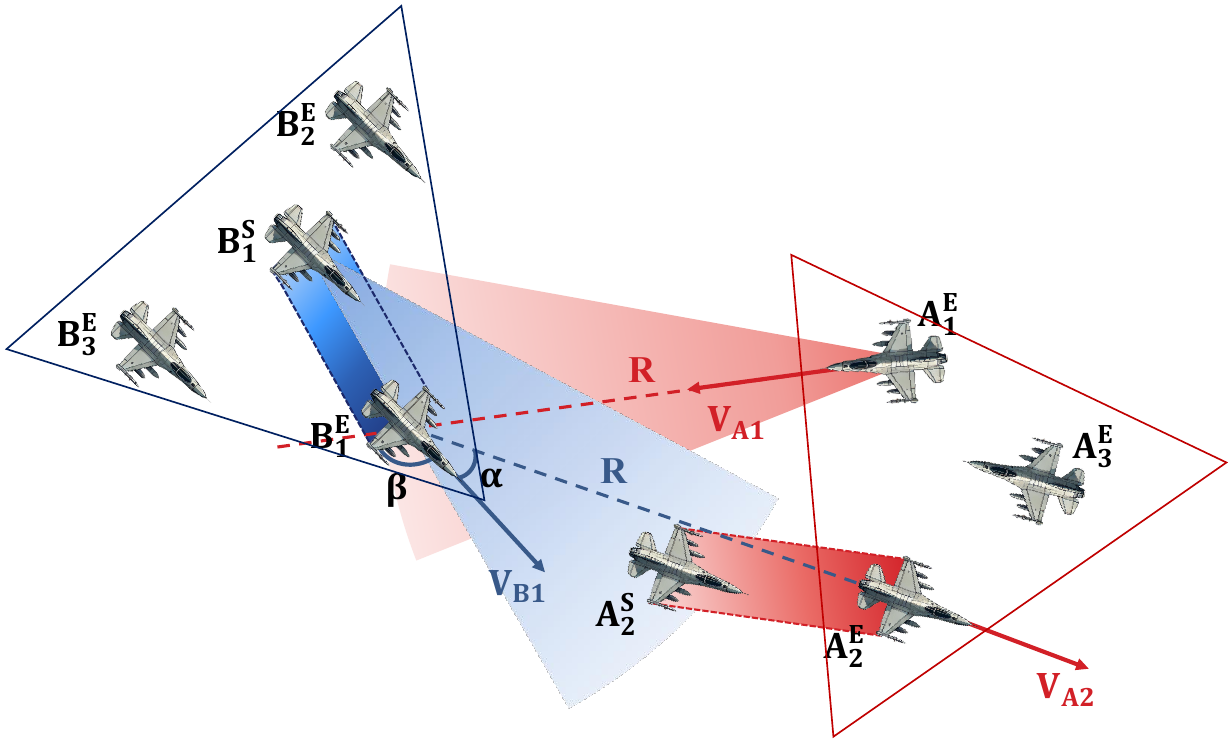} 
\caption{Schematic diagram of multi-UAV combat mission. The UAVs are divided into two teams. The superscripts S and E represent the starting and ending positions. The dashed lines show the relative distances between the UAVs, and the vector arrows indicate their velocity directions.
Initially, $A_2$ (follower) is positioned within the weapon engagement zone (WEZ) \cite{dillon2023optimal} of $B_1$, represented by the sector-shaped area. $A_2$ accelerates to escape from $B_1$'s WEZ, and as $B_1$ pursues, it enters the WEZ of $A_1$ (leader).
}
\label{fig:task}
\end{figure}

While previous studies have made significant advancements in UAV maneuvering and coordination, many existing approaches face challenges in handling high-dimensional state and action spaces and ensuring effective collaboration in dynamic multi-UAV combat environments.  
In this paper, the UAVs are divided into two teams, with each team comprising leaders and followers. The leader manages the overall strategy, while the followers closely follow the leader’s guidance. Figure~\ref{fig:task} illustrates a bait-and-strike combat strategy. 
We focus on addressing the high-dimensional and collaborative challenges of UAVs. We propose a hierarchical framework with a leader-follower strategy, which has been validated in a simulation environment.  
The contributions of this paper are as follows:  
\begin{itemize}
	\item 
	We propose a hierarchical framework to address the high-dimensional challenges.  
The framework consists of three levels. A top level that conducts macro-level analysis of the combat situation, a middle level that generates desired action angles for UAV agents, and a bottom level that translates these angles into specific action commands.
	\item 
	 To enhance coordination among UAVs, we develop a top-level policy selector for the leader-follower feature. This policy optimizes the agents' approximate state-value functions by assigning distinct roles to the UAVs, where followers estimate the leader’s utility, improving cooperative behaviors and overall mission performance.  
   \item 
   We introduce a target selector that evaluates the threat level of different targets based on multiple dimensions, such as flight status and posture, ensuring more precise and effective target selection.
	\item 
 Extensive simulation experiments validate the effectiveness of the proposed framework. Compared with traditional methods, our approach can obtain higher reward values and better trajectories.
\end{itemize}
\section{Related work}
\subsection{Hierarchical reinforcement learning}
Hierarchical Reinforcement Learning (HRL)~\cite{pateria2021hierarchical} addresses the inefficient learning and the convergence issues caused by high-dimensional state and action spaces. 
By introducing multiple levels of abstraction, HRL enables the decomposition of complex tasks into simpler and more distinguishable sub-tasks~\cite{li2024interactive}, reducing the complexity of the problem space and enhancing the learning process by facilitating more efficient and targeted decision-making at various levels.
Chen et al.~\cite{chen2023hierarchical} utilized hierarchical reinforcement learning (HRL) to transform a three-dimensional problem into a two-dimensional one, effectively mitigating the non-convergence issues associated with high-dimensional complexity. This strategy reduces the complexity of training through dimensionality reduction.
Pope et al.~\cite{pope2021hierarchical,pope2022hierarchical} proposed a hierarchical structure that divides tasks into sub-tasks, solving high-dimensional continuous control problems in 1v1 air combat scenarios.
Kong et al.\cite{kong2023hierarchical} extended the application to variable-scale formation air combat tasks by designing three sub-tasks and employing a cyclic soft actor-critic algorithm to learn sub-policies, enabling formations to exhibit effective cooperative behavior in both symmetric and asymmetric environments.
To further improve the functionality of hierarchical strategies, recent studies have introduced explicit task division within hierarchical architectures. Mei et al.~\cite{mei2024deep} proposed a two-layer decision-making framework in which the high-level strategy generates maneuvering commands under combat conditions, while the low-level strategy computes specific control signals for aircraft. Similarly, Chai et al.~\cite{chai2023hierarchical} separates the problem into macro and micro perspectives, generating corresponding actions at each level to enhance control efficiency.
By creating an effective link between strategic decision-making and control execution, these architectures provide more flexible solutions for complex tasks.
\subsection{Leader-follower formation control}
The leader-follower formation represents an efficient organizational approach for team coordination and combat operations in dynamic and complex battlefield environments. The formation inherently establishes a hierarchical structure that enables leaders to assume a directive role to provide strategic guidance, while followers execute tasks aligned with the leader's intent, thereby enhancing the overall effectiveness of coordinated actions. 
Wang et al.~\cite{wang2020leader} proposed a leader-follower-based cooperative formation trajectory tracking control method for multi-UAV systems, employing an integral backstepping approach to design the leader UAV's trajectory controller and a sliding mode controller to achieve precise formation control. Building on this foundation, Hao et al.~\cite{hao2021formation} introduced a distributed leader-follower formation control approach that aligns the follower UAVs' headings with that of their leader while maintaining the desired relative positions, enabling a more flexible distributed cooperative control mechanism. Ranjan et al.~\cite{ranjan2024relational} proposed a novel leader-follower formation control method from a behavior modeling perspective, defining the follower UAVs’ linear and angular velocities as control inputs by simulating human pilot behaviors.
To enhance the intelligence of leader-follower strategies, Liu et al.~\cite{liu2024leader} integrated Deep Q-Networks into leader-follower formation control, enabling adaptive adjustments to formation behaviors. Wang et al.~\cite{wang2024hierarchical} developed a leader-follower PID-based formation control approach that incorporates a virtual-structure-based reinforcement learning scheme to construct a hierarchical multi-agent training framework, achieving precise formation-level control.
Existing leader-follower approaches primarily focus on maintaining formation stability by coordinating the positions of leaders and followers, with followers maintaining relatively fixed positions relative to the leader. However, these methods often overlook the critical aspect of follower coordination during combat missions. In this paper, we emphasize the leader-follower relationship and propose an approach to enhance UAV cooperation in combat scenarios.
\section{Background}
\subsection{Problem description}
In multi-UAV combat, two opposing forces, designated as Blue and Red, aim to destroy the UAVs of the opposing team while minimizing damage to their own side inflicted by the opponent.
In multi-UAV combat, two opposing forces, designated as Blue and Red, aim to destroy the UAVs of the opposing team while minimizing damage to their own side inflicted by the opponent.
Modern fighter operations utilize a two-UAV or three-UAV formation combat mode, with the leader serving as the primary and the follower providing auxiliary support.
In this study, referencing this formation, heterogeneous UAVs with varying combat capabilities are categorized as leaders and followers. Each UAV is equipped with radar and other sensors to gather information about both allies and opponents.
We employ the six-degree-of-freedom flight dynamics model provided by JSBSim~\cite{vogeltanz2016survey} to simulate UAV movement within the environment. This model is extensively utilized in UAV research and serves as a foundational tool for accurately representing UAV flight dynamics.
\subsection{State space}
The states of UAVs include their own states and relative states with respect to other UAVs. We introduce a set of symbols and definitions related to aircraft flight posture and control to accurately model the movement.
To represent the UAV's state, an 11-dimensional vector 
$s$ is used, as shown in Table~\ref{tab:state}. The vector $s$ includes position, orientation, velocity, and other relevant parameters that influence decision-making.
In multi-UAV air combat, to enhance each UAV’s situational awareness of the environment, it is crucial for each UAV to consider the spatial relationships between allies and targets when making maneuver decisions. 
These relationships are represented using relative states,  including the angle off $\alpha$ and target angle $\beta$.
The angle off $\alpha$ is primarily used for navigation and path planning. Smaller values of $\alpha$ indicate that the UAV is more closely aligned with the target. This alignment reduces the complexity of the path planning and allows the UAV to focus on optimal routes toward the target.
The target angle $\beta$ describes the specific 
location of the target within the UAV's field of view. Smaller values of $\beta$ indicate that the target is directly in front of the UAV or very close to the forward direction, making it easier for the UAV to track and engage the target effectively.
\begin{table}[width=.9\linewidth,cols=3,pos=h]
\caption{Variables and ranges of the UAV state.}\label{tab:state}
\begin{tabular*}{\tblwidth}{@{} LLL@{} }
\toprule
\textit{Symbol} & \textit{Range} & \textit{Description} \\
\midrule
$p_x, p_y, p_z$ & [0, 1000m] & The coordinates of UAV in the environment \\
$v_x, v_y, v_z$ & [-50m/s, 50m/s] & Velocity components of the UAV along three axes \\
$\phi$ & [$-\pi$, $\pi$] & Roll angles of the UAV \\
$\theta$ & [$-\pi$, $\pi$] & Pitch angles of the UAV \\
$\psi$ & [$-\pi$, $\pi$] & Yaw angles of the UAV \\
$\alpha$ & [$-\pi$, $\pi$] & Angle between UAV’s current heading and the direction to target \\
$\beta$ & [0, $2\pi$] & Desired angle that UAV needs to achieve to align with target \\
\bottomrule
\end{tabular*}
\end{table}

\subsection{Action space}
The six-degree-of-freedom (6-DOF) model provides a more comprehensive description of UAV movement compared to the 3-DOF model, enabling maneuverability in three-dimensional space. It allows the UAV to translate along the $x$, $y$, and $z$ axes for precise positioning and to rotate around these axes to adjust its roll $\phi$, pitch $\theta$, and yaw $\psi$ angles in the environment. These rotations enable the UAV to adjust its posture and direction effectively.
By setting the rotational motions $\delta_\phi$, $\delta_\theta$, and $\delta_\psi$ along the three axes, as well as adjusting the throttle $\delta_{th}$, the aircraft's direction and speed can be precisely controlled. This combination of translational and rotational movements enables the UAV to execute complex maneuvers. Consequently, the action set for the aircraft is defined as $a = \{\delta_\phi, \delta_\theta, \delta_\psi, \delta_{th}\}$, encompassing all necessary control inputs for effective movement within its environment.
\subsection{Reward function}
The reward function quantitatively measures the desirability of a state-action pair, guiding the UAV toward achieving its objectives. Positive rewards are typically assigned to desirable outcomes, while negative rewards correspond to undesirable ones.
It serves as a critical feedback mechanism, informing the UAV's decision-making process and ensuring alignment with mission goals.
In our framework, the reward function consists of posture reward, distance reward, and event reward components.
The event reward is influenced by the weapon's attack zone, which restricts the UAV's ability to strike targets. The attack zone is defined as a sector with a radius of 4 km and an angle of 45 degrees.
\section{Methodology}
\begin{figure}[]
  \centering
  \includegraphics[width=\linewidth]{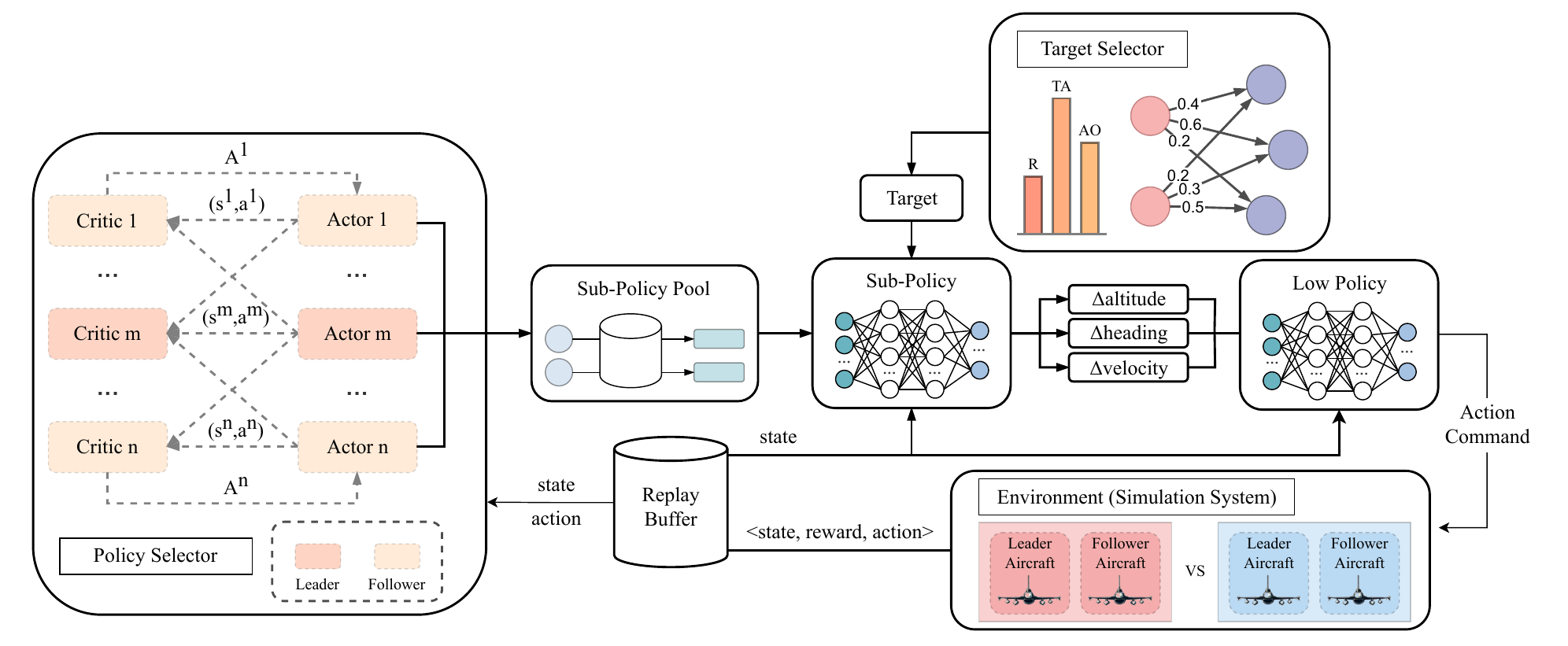}
  \caption{Overview of the Proposed 
 Hierarchical Leader-Follower control framework.
  The top level, a policy selector oversees macro-level decision-making, assigning appropriate sub-policies based on different UAV tactical strategies. The middle level generates desired action adjustments in response to the current state, while the bottom level translates these into specific action commands. A target selection module provides the middle level with target information to enhance decision-making efficiency.}
  \label{fig:framework}
\end{figure}
\subsection{Cooperative hierarchical framework}
As the number of UAVs in air combat missions and the dimensionality of state variables in 6-DOF model increase, the high-dimensional state and action spaces pose significant challenges to the training process.
Traditional algorithms typically decompose many-on-many air combat missions into multiple one-on-one engagements, which fail to adequately simulate collaboration among UAVs in real combat scenarios. To address these challenges, we propose a collaborative hierarchical framework that divides the decision-making process into three layers, as shown in Figure~\ref{fig:framework}.

\subsubsection{Policy selector with leader-follower strategy}
The policy selector, as the global controller, evaluates and selects suitable sub-policies for different scenarios, enabling effective collaboration among UAVs and ensuring the achievement of overall mission objectives.
In a one-on-one UAV battlefield, a UAV only needs to consider the opponent's policy to make decisions. However, collaboration among UAVs is a key factor in achieving combat effectiveness beyond that of a single UAV in multi-UAV combat missions. Existing works simplify multi-UAV combat into multiple one-on-one battles, neglecting the relationships among allies and failing to enable effective cooperation.
To address this limitation, we propose a cooperative policy selector utilizing a leader-follower strategy. In the framework, the follower considers the leader's policy when choosing an action, enabling  collaboration between the leader and the follower.
The effectiveness of collaboration is reflected through rewards received from the environment based on the agents' states and actions. Reinforcement learning aims to maximize cumulative rewards discounted over time. The return of an agent is defined as:
\begin{equation}
    G_{t} = \sum_{k=0}^\infty \gamma^k r_{t+k+1} ,
\end{equation}
where $\gamma \in \left[0, 1 \right]$ is the discount factor.

Designed for cooperative and competitive scenarios, MAPPO is a robust multi-agent reinforcement learning algorithm. It leverages the classical actor-critic framework and employs a neural network to approximate both the policy and the value function.
The actor $\pi(s; \theta_{a})$ maps states to action distribution probabilities, enabling agents to make stochastic decisions. The critic $V(s; \theta_{c})$ approximates the state value function $V(s) = \mathbb{E}[G | s_0 = s]$, which evaluates the quality of the policy.
Agents interact with the environment to collect multiple trajectories, which are stored in an experience buffer. At each step, an agent selects an action based on the current policy and receives the next state and reward from the environment through state transitions.
The advantage function is calculated using generalized advantage estimation (GAE), defined as:
\begin{equation}
    A_{t} = \sum_{l = 0}^{k}(\gamma \lambda)^{l} \delta_{t+l}   ,
\end{equation}
where $\delta_{t} = r_{t}+\gamma V_{\theta_{c}}\left(s_{t+1}\right)-V_{\theta_{c}}\left(s_{t}\right)$ and $A_{t}$ is the advantage of taking action $a_t$ in state $s_t$.
The critic is a centralized value function that evaluates the joint states and actions of all agents from a global perspective to accurately assess policies.
However, using the same value function for all agents fails to capture the differences in roles among them.
To address this issue, we propose an improvement to the critic that influences action selection. In our framework, the leader focuses on global information, while the follower devotes more targeted attention to the leader's policy when determining actions.Even when the leader's action is suboptimal, the follower is able to achieve higher rewards.
To further enhance collaboration, a max-min criterion is employed to optimize the policy, maximizing the minimum possible outcome. The leader-follower update value functions as follows:
\begin{equation}
\begin{aligned}
 V_f(x_t^f) = & \left(1-\alpha\right)V_f(x_t^f) 
  +  \alpha \left[r_t^f + \gamma \mathop{\max}_{a^f \in A^f} \mathop{\min}_{a^l \in A^l} V_f(x_{t+1}^f)\right] ,\\
  V_l(s_t^l) = & \left(1-\alpha\right)V_f(s_t^l) + \alpha \left[r_t^l + \gamma V_l(s_{t+1}^l)\right] ,
\end{aligned}
\end{equation}
where $V_f$ is the follower's value function, $V_l$ is the leader's value function, and 
$x_t^f=\left(s_t^f, a^l \right)$ is the follower's state. 
Optimizing critic using the stored data from the experience buffer with a loss function of:
\begin{equation}
    L_{critic}(\theta_c) = \mathbb{E}_{\left(s_{t}, a_{t}\right) \sim \mathcal{D}}\left[V\left(s_{t} ; \theta_c\right)-G_{t}\right]^{2} .
\end{equation}
At each time step, the expected advantage function is maximized by maximizing the policy loss objective function:
\begin{equation}
\begin{aligned}
    L_{actor }(\theta_a) =  & \mathbb{E}_{\left(s_{t}, a_{t}\right) \sim \mathcal{D}}
    \left[ \min 
        \left(\rho\left(s_{t}, a_{t} ; \theta_{a}\right) A_{t},\right.\right. 
\left.\left.\operatorname{clip} \left[\rho \left(s_{t}, a_{t} ; \theta_{a}\right), 1-\epsilon, 1+\epsilon\right] A_{t}\right)\right] ,\\
     \quad 
    \rho\left(s, a;\theta_{a}\right)  = & \frac{\pi(a|s;\theta_a  )}{\pi_{old}(a|s;\theta_a)} ,
\end{aligned} 
\end{equation}
where $\pi$ is the current policy, $\pi_{old}$ is the old policy, $\epsilon$ is a hyperparameter that controls the clipping range.
The combination of policy loss and value loss also includes an entropy reward to encourage exploration, so that the overall objective function is:
\begin{equation}
    L = L_{critic}(\theta_c) + L_{actor}(\theta_a) + \mathcal{H} ,
\end{equation}
where $\mathcal{H}$ is the policy entropy averaged over all agents’ entropy $\mathcal{H}_i=- \sum_{a} \pi(a|s) \log \pi(a|s) $.

\subsubsection{sub-policy}
Different sub-policies are selected by the policy selector to execute various flight maneuvers based on the current states. Training these sub-policies relies on the movement substructure of the hierarchical framework, which encompasses the middle and bottom layers. The movement substructure addresses various combat scenarios by training a series of sub-policies that can handle the complex dynamics of specific tactical situations and provide flexible responses, enabling diverse combat options.
We categorized combat policies into three types:
\begin{enumerate}[\textbullet]
\item The Target-Approaching Policy: Aims to minimize the distance to enemies to enable effective attacks or surveillance.
\item The Offensive Policy: Focuses on proactively attacking enemies to gain an advantage.
\item The Defensive Policy: Ensures safety by evading enemy attacks while seeking opportunities to counterattack.
\end{enumerate}
The sub-policy flight problem is modeled as a Markov Decision Process (MDP)~\cite{zhong2024theoretical}, defined as $G=\{S, A, R, T \}$, where $S$ is state space, $A$ is action space, $R$ is reward function, and T is state transition function. The state consists of the aircraft state and the relative state.
The aircraft state includes position, posture, and altitude, while the relative state is characterized by the angle, as shown in Figure~\ref{fig:task}. The UAV evaluates its posture based on azimuth and track angle and controls roll, yaw, pitch, and throttle through commands to the aileron $\delta_\phi$, elevator $\delta_\theta$, rudder $\delta_\psi$ and throttle $\delta_{th}$.

We propose a reward function to quantify the posture and distance-based threat relative to enemies. The posture reward evaluates the flight posture and orientation of the aircraft, emphasizing favorable flight conditions that lead to higher rewards. It is defined as follows:
\begin{equation}
\begin{aligned}
      r_{a} = &
      \min \left(\operatorname{arctanh}\left(1-\max \left(\frac{\alpha+\beta}{\pi}, 10^{-4}\right)\right)/ \pi, 0\right) 
        +  \frac{2\pi}{25 \cdot\left(\alpha+\beta \right) +2 \pi}
      +\frac{1}{2}.
\end{aligned}
\end{equation}
The distance reward component encourages proactive engagement while discouraging passive or evasive maneuvers by the aircraft. It is formulated as:
\begin{equation}
    r_{d} = 
    \begin{cases} 
    \frac{e^{k_1 x}}{e^{k_1 x} + \frac{1}{x}} + b_1, & \text{if } x < \frac{1}{3}, \\
    x, & \text{if } \frac{1}{3} \leq x < \frac{2}{3}, \\
    -\frac{e^{k_2 x + b_2}}{b_3} + b_4, & \text{if } x \geq \frac{2}{3},
    \end{cases} 
\end{equation}
where $x$ represents the ratio of the current distance 
$d$ to the maximum attack range 
$D_{\max}$.
This motivates the aircraft to maintain proximity to strategic targets or optimal engagement zones, thereby enhancing its operational effectiveness.
The state transition is determined by the flight control module, relying on the JSBSim simulation system. The system generates the next state by $s'=f(s, g(\delta^a_\theta))$, where $\delta^a_\theta$ is the target angle to realize the state transition. Actions are input into the environment at a maximum simulation frequency of 50 Hz, ensuring real-time adaptability to dynamic conditions.

Various sub-policies are trained by setting distinct initial state ranges and respective reward functions.
In addition, due to the functional differences between the leader and the follower, the follower is able to effectively lure enemies and assist the leader in accomplishing combat missions via more aggressive and more flexible flight policies at a lower cost advantage.

\subsection{Target Selector}
In multi-UAV combat missions, simply selecting nearby targets fails to adequately assess the true threat posed by each UAV. Conversely, situational target selection offers numerous advantages. By employing situational target selection, global tactics and strategies can be comprehensively optimized, thereby mitigating the risk of UAV losses and ensuring overall operational success.
The angular relationship between a UAV and its target is pivotal in the target selection process. While traditional algorithms typically consider the impact of angle off and target angle on the situation, these metrics may sometimes fall short in accurately reflecting the UAV's actual threat level. To address this, we propose employing the n-step method to more precisely gauge UAV threat levels, factoring in the dynamics and evolving patterns of their situational awareness.

Intuitively, the closer the target is, the easier it is for the aircraft to approach and attack. The threat level of the aircraft is fully considered in the target scoring, including the position attribute $T_d$ and the posture attribute $T_a$. The posture reflects the potential threat level of the target to the aircraft and the mission. Targets with a higher threat level need to be handled first. Due to the varying functional attributes of different aircraft, it is crucial to evaluate the ability level of the target accordingly. This formula ensures a balanced consideration of critical factors:
\begin{equation}
    S = \alpha T_d + \beta T_a + \gamma I ,
\end{equation}
where the coefficients $\alpha$, $\beta$ and $\gamma$ are weighting factors that adjust the importance of each attribute based on the mission requirements and aircraft capabilities.

\section{Experiment}
\subsection{Experimental Setting}
In this paper, we select F-16 combat aircraft as agents for simulation. The agents are divided into two teams, red and blue, with an equal number of agents on each side. The parameters of the air combat environment are configured as follows: the maximum missile interception range is 4 kilometers, and the field of view is $\pi$/4. In the aircraft motion model, the maximum axial acceleration is set to 10 $m/s^2$, and the minimum flight altitude is restricted to 3 kilometers. Each team consists of six aircraft, with their initial positions randomly assigned. The control signal is normalized to [-1,1], enabling smooth transitions in the action space. The aircraft adjust their motion posture in the JSBSim simulation environment via action commands.
A UAV is considered destroyed if its altitude drops below the minimum threshold, exceeds overload limits, collides with another UAV, or is hit by a missile. A missile strikes a UAV if it is within the missile's field of view and the distance is less than 300 meters.
The parameter settings for the experiments are as follows: the Proximal Policy Optimization (PPO) clip is set to 0.2, the Generalized Advantage Estimation (GAE) $\lambda$ is 0.95, and the buffer size is 3000. The discount factor is 0.99, the activation function is ReLU, and the optimizer is Adam with a learning rate of $3e^{-4}$.
To evaluate the efficiency of the model, we compare our method with the following baseline approaches:
(a) \textit{MAPPO}~\cite{yu2022surprising}  incorporates the advantages of PPO into multi-agent environments, 
demonstrating superior performance in addressing multi-agent cooperation and competitive tasks.
(b) \textit{PPO}~\cite{schulman2017proximal} introduces a clipping mechanism to restrict policy updates and optimizes them with the advantage function.
(c) \textit{VDN}~\cite{sunehag2017value} decomposes the global joint value function into the sum of the local value functions for each agent. 
(d) \textit{QMIX}~\cite{rashid2020monotonic} extends VDN by employing a hybrid network to capture the nonlinear cooperative relationships between agents.
For a fair comparison, all methods utilize the same states, actions, and reward structures during evaluation.

\subsection{Results}
The following experiments are conducted to demonstrate the superior effectiveness of the LFMAPPO algorithm compared to other methods in executing multi-UAV combat tasks.

\textbf{Evaluation criteria.} 
The evaluation is conducted in two principal dimensions: reward and win rate.
The process dimension focuses on the reward to evaluate the system's performance during task execution, while the outcome dimension uses the win rate to assess the mission's overall success.

\begin{figure}[htbp]
    \centering
    \begin{subfigure}[b]{0.45\linewidth}
        \centering
        \includegraphics[width=0.8\linewidth]{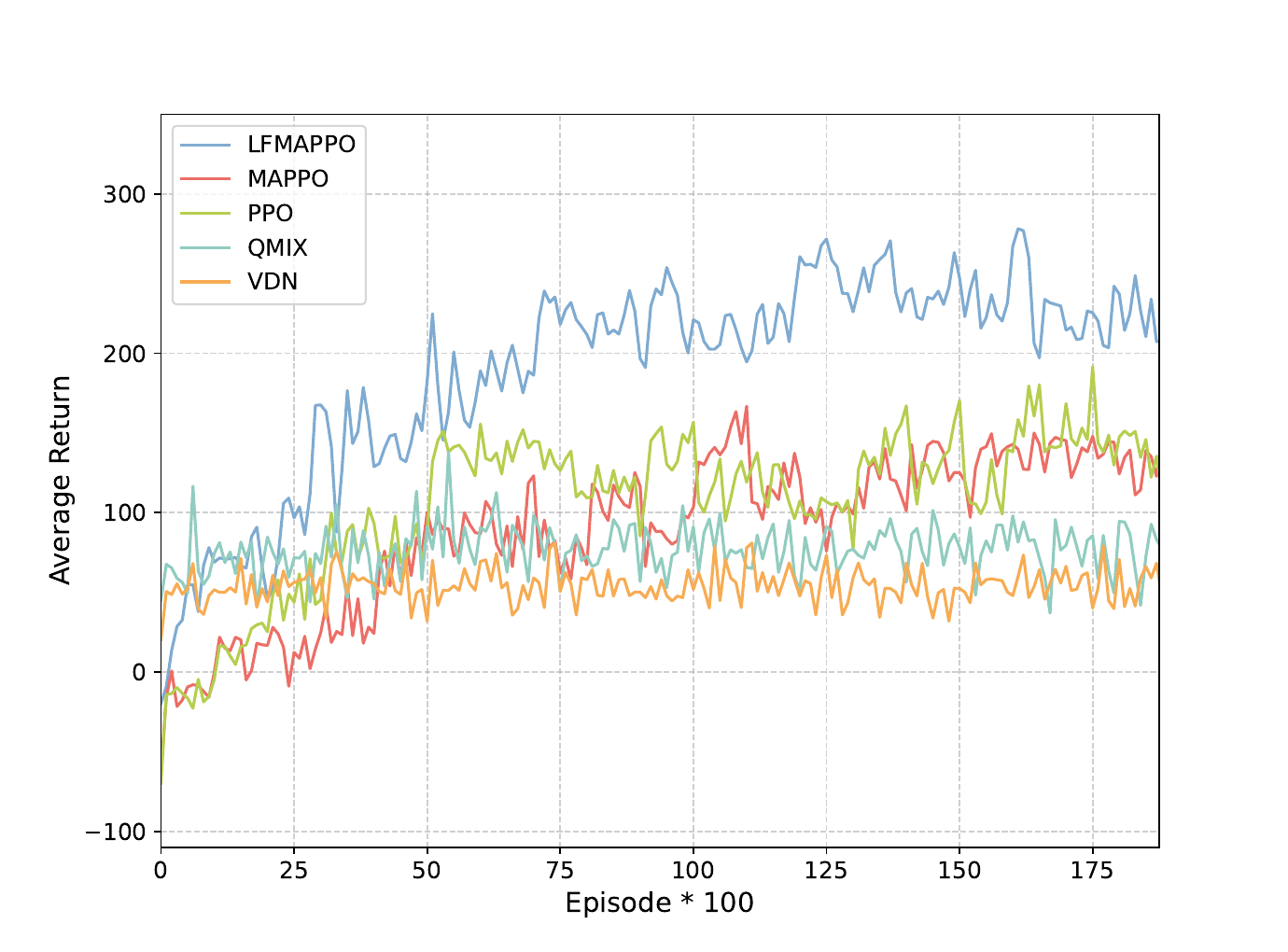}
        \caption{Average return curve}
        \label{fig:exp:main-reward}
    \end{subfigure}
    \hfill
    \begin{subfigure}[b]{0.45\linewidth}
        \centering
        \includegraphics[width=0.9\linewidth]{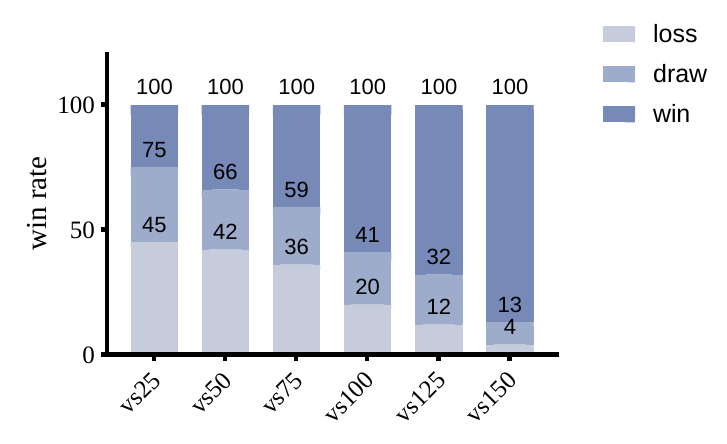}
        \caption{Win-Draw-Loss trends}
        \label{fig:exp:main-win}
    \end{subfigure}
    \caption{\textbf{The effectiveness of the algorithm is analyzed for two perspectives, average reward and win rate.}
(a) Average return curve: The average reward values of different methods during UAV training tasks over training episodes. As training progresses, the performance of each method improves gradually. 
Our approach  demonstrate superior performance in the task, achieving higher reward values.
(b) Win-Draw-Loss trends: The trends in win rates, draw rates, and loss rates of UAVs across different episodes. 
The win rate gradually increases, while the draw and loss rates decline correspondingly, reflecting the progressive improvement of the model.
}
    \label{fig:exp}
\end{figure}

\textbf{Main results.} 
Figure~\ref{fig:exp:main-reward} illustrates the average return achieved by our approach and baseline algorithms during combat against an opposing team, with each method evaluated over five independent runs. Our approach consistently demonstrates superior performance, achieving higher rewards throughout the training episodes compared to the baseline methods.
In contrast, VDN and QMIX exhibit limited learning capabilities in this environment, primarily due to their discretization of the action space, which reduces movement flexibility and limits strategic diversity. Furthermore, their inability to handle high-dimensional states imposed by six-degree-of-freedom motion further contributes to instability during the training process.
The higher rewards achieved by our approach can be attributed to its evaluation of policies using a value function. By utilizing the leader-follower framework, it ensures that different roles receive clear and distinct rewards, forming well-defined objectives: leaders make decisions based on their local observations, while followers act based on the leaders' decisions. This cooperative dynamic enables the algorithm to effectively leverage the local state structure, enhancing teamwork and coordination. As a result, the agents adapt more effectively to dynamic combat scenarios, consistently surpassing the baselines in overall reward accumulation.
 
\begin{figure}[htbp]
    \centering
    \begin{subfigure}[b]{0.45\linewidth}
        \centering
        \includegraphics[width=0.8\linewidth]{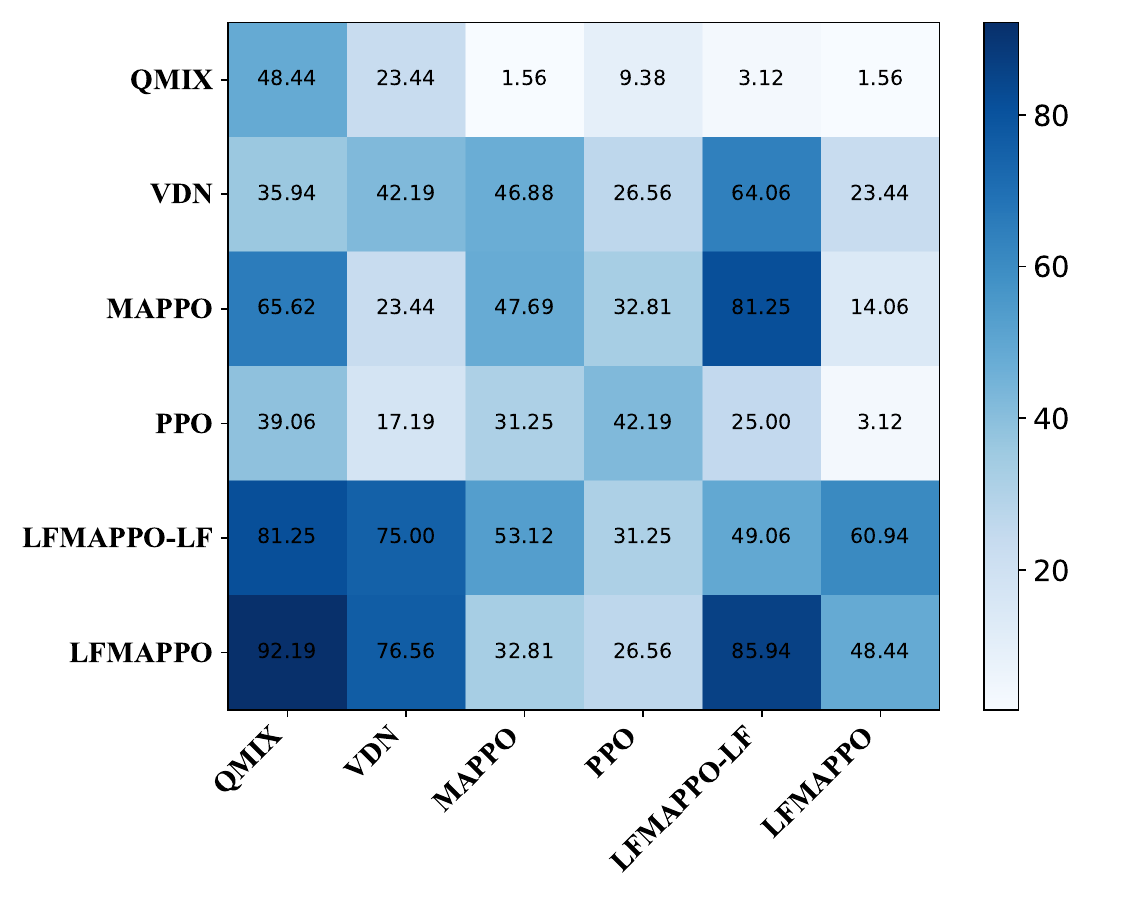}
        \caption{50 episodes}
        \label{fig:exp:win-50}
    \end{subfigure}
    \hfill
    \begin{subfigure}[b]{0.45\linewidth}
        \centering
        \includegraphics[width=0.8\linewidth]{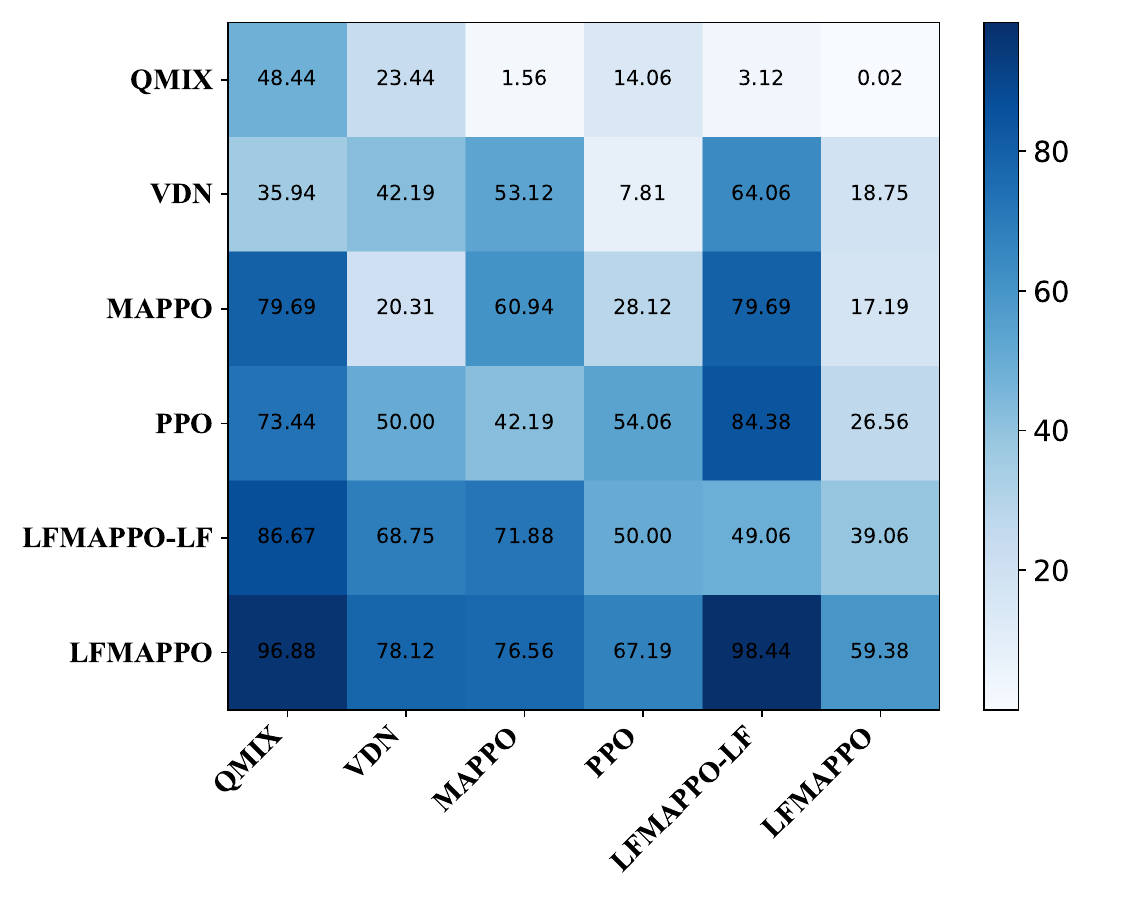}
        \caption{100 episodes}
        \label{fig:exp:win-100}
    \end{subfigure}
    \caption{
    \textbf{A comprehensive comparative analysis of algorithms in the air combat environment.} It shows respective win rates after training for 50 episodes (a) and 100 episodes (b). please check grammar
    }
    \label{fig:win}
\end{figure}

To evaluate the effectiveness of our method, we test the policies at various stages of training against a baseline opponent trained for 100 episodes, recording the win, draw, and loss rates at each stage.
A total of 128 combats are conducted per stage, with the initial state of each combat randomized.
The results, shown in Fig.~\ref{fig:exp:main-win}, indicate that as the number of iterations increases, the draw rate decreases, while a notable upward trend in the win rate shows the system's improved ability to defeat opponents. 

In multi-UAV combat scenarios, we randomly generated the initial states of the UAVs within a specified range. Figure~\ref{fig:win} illustrates the win rate results from a 128-episode cross-comparative experiment, showing that the LFMAPPO algorithm outperforms other methods, demonstrating the feasibility of our proposed approach. Figure~\ref{fig:exp:win-50} and ~\ref{fig:exp:win-100} depict the win rates after 50 and 100 episodes of training, respectively. A comparison between these figures reveals a general increase in win rates as the number of training episodes grows. 

\begin{figure}[htbp]
    \centering
    \begin{subfigure}[b]{0.45\linewidth}
        \centering
        \includegraphics[width=0.85\linewidth]{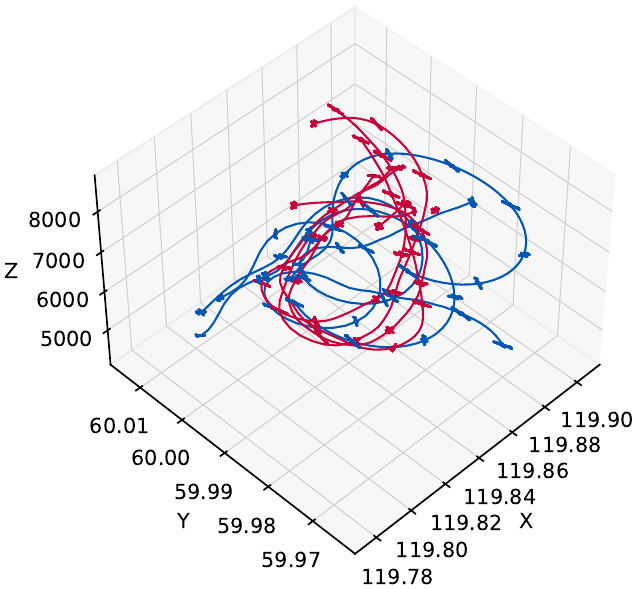}
        \caption{MAPPO~\cite{yu2022surprising}}
        \label{fig:exp:traj-mappo}
    \end{subfigure}
    \hfill
    \begin{subfigure}[b]{0.45\linewidth}
        \centering
        \includegraphics[width=0.85\linewidth]{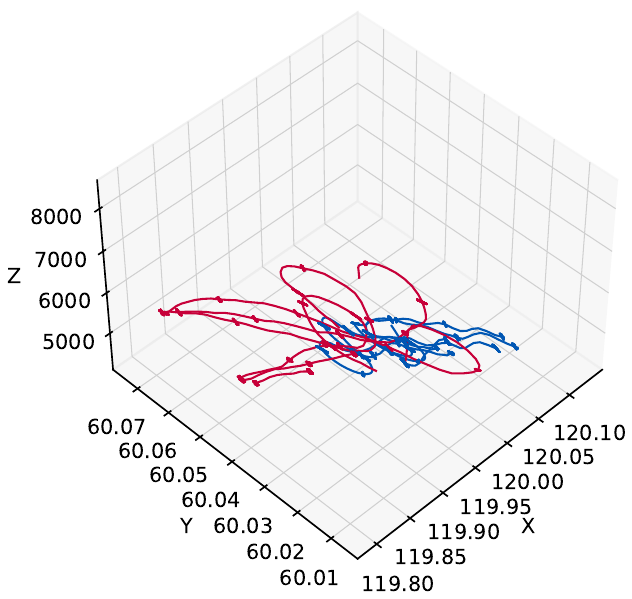}
        \caption{PPO~\cite{schulman2017proximal}}
        \label{fig:exp:traj-ppo}
    \end{subfigure}
    \vspace{1em} 
    \begin{subfigure}[b]{0.45\linewidth}
        \centering
        \includegraphics[width=0.85\linewidth]{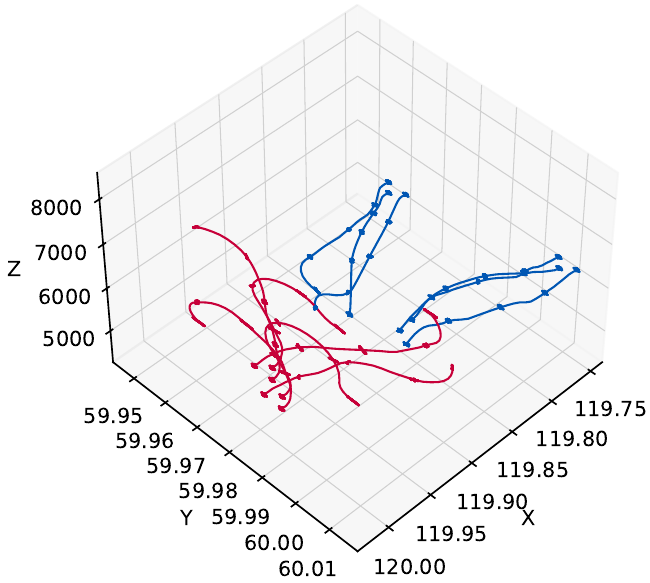}
        \caption{LFMAPPO (Combat dynamic)}
        \label{fig:exp:traj1}
    \end{subfigure}
    \hfill
    \begin{subfigure}[b]{0.45\linewidth}
        \centering
        \includegraphics[width=0.85\linewidth]{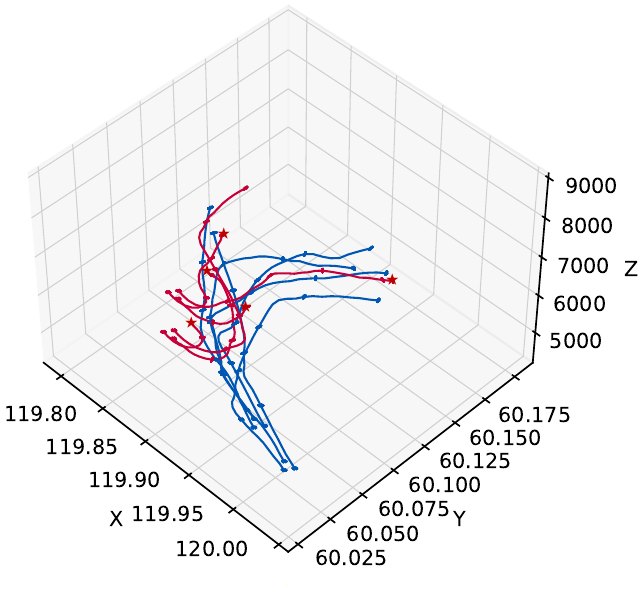}
        \caption{LFMAPPO (Maneuver trajectory)}
        \label{fig:exp:traj4}
    \end{subfigure}
    \vspace{1em} 
    \begin{subfigure}[b]{0.45\linewidth}
        \centering
        \includegraphics[width=0.85\linewidth]{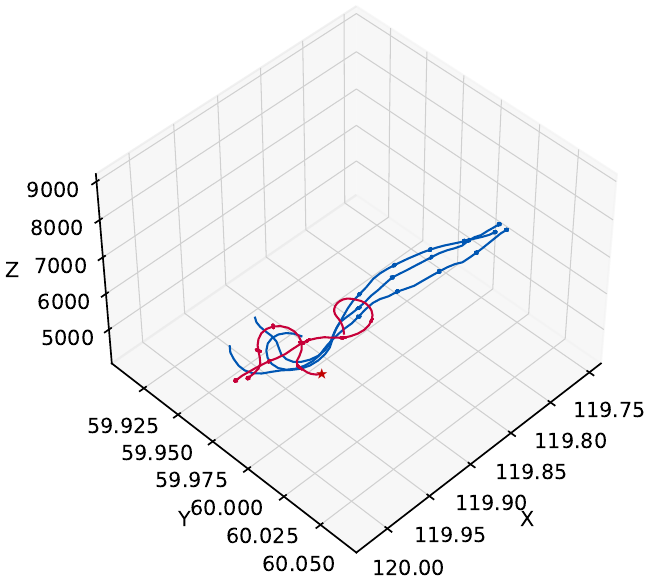}
        \caption{LFMAPPO (Pursuit trajectory)}
        \label{fig:exp:traj2}
    \end{subfigure}
    \hfill
    \begin{subfigure}[b]{0.45\linewidth}
        \centering
        \includegraphics[width=0.85\linewidth]{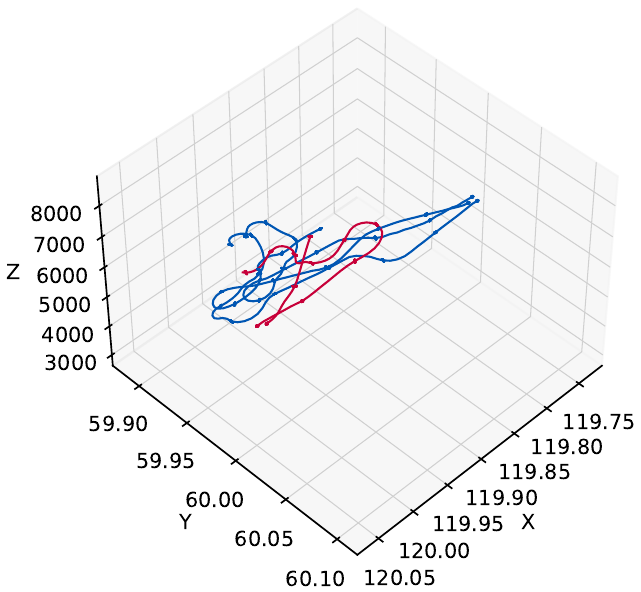}
        \caption{LFMAPPO (Bait-strike trajectory)}
        \label{fig:exp:traj3}
    \end{subfigure}
    \caption{\textbf{The maneuver trajectories of different algorithms in a 6v6 multi-UAV combat scenario.} (a), (b) and (c)-(f) represent the UAV maneuver trajectories using the MAPPO, PPO, and LFMAPPO algorithms, respectively. (c) and (d) illustrate the flight dynamics of both sides in a three-dimensional combat scenario. (e) and (f) present segments of the complete UAV combat trajectories, illustrating the continuous pursuit and bait-strike attack strategies.}
    \label{fig:exp:1}
\end{figure}
\textbf{Behavior analysis.} 
We plotted the combat trajectories in a multi-UAV scenario to analyze the behavior of the aircraft. As shown in Figure~\ref{fig:exp:1}, the red trajectories represent the opposing force. The x and y coordinates correspond to latitude and longitude and z represents the altitude.
Figure~\ref{fig:exp:traj-mappo} and~\ref{fig:exp:traj-ppo} illustrate the trajectories of the MAPPO and PPO algorithms in a three-dimensional combat scenario. Compared to the trajectories of the LFMAPPO algorithm, these methods commonly form looping patterns, with UAVs circling around each other. This behavior extends the task execution time and hinders the ability to quickly neutralize opponents. Moreover, during these prolonged circling engagements, agents accumulate reward values, which inflate the reward scores for these methods. However, these scores are deemed ineffective as they do not contribute to achieving mission targets. In contrast, the maneuver trajectories produced by our approach enable rapid and decisive attacks, allowing UAVs to efficiently defeat opponents.
In Figure~\ref{fig:exp:traj1} and~\ref{fig:exp:traj4}, we observe that UAVs form cooperative maneuver trajectories as groups.

\textbf{Effect of policy selection with the leader-follower strategy.}
The blue trajectories, guided by a leader-follower strategy, maintain formation as they to approach the target and launch coordinated attacks, as shown in Figure~\ref{fig:exp:traj1}-~\ref{fig:exp:traj3}.
In Figure~\ref{fig:exp:traj1}, the movement trajectories of blue aircraft and red aircraft in three-dimensional space exhibit distinct differences. The red aircraft's trajectories appear scattered and lack a cohesive tactical formation. In contrast, the blue aircraft, advancing in groups of three, display a well-coordinated movement pattern as they approach the mission target.
In Figure~\ref{fig:exp:traj4}, the blue aircraft demonstrated effective teamwork by cooperatively engaging and successfully defeating the red aircraft. Initially, the red aircraft adopted a defensive posture within the weapon’s engagement zone, maneuvering to evade potential threats.
Despite employing evasive tactics, such as sharp turns and acceleration to escape the engagement zone, the red aircraft was pursued. The blue aircraft accelerated their pursuit, synchronizing their movements to close the distance. The trajectories of the blue aircraft illustrate their coordination and effective teamwork.
The red aircraft executed continuous attacks, as shown in Figure~\ref{fig:exp:traj2}. The blue aircraft first cooperated to attack the red aircraft, while another red aircraft pursued the blue aircraft. After successfully downing one red aircraft, the blue aircraft shifted focus to another target and initiated an attack. The blue aircraft accelerated and adjusted their flight paths to engage the red aircraft pursuing them.
Throughout the engagement, the red aircraft maintained a triangular formation, serving as a stable and effective combat structure in a multi-agent scenario. This formation allowed the red aircraft to coordinate their attacks effectively, providing both offensive and defensive advantages and maximizing the potential for strategic maneuvering.
In Figure~\ref{fig:exp:traj3}, the blue aircraft executed a bait-and-strike strategy. One of the blue aircraft accelerated along a relatively direct flight path, serving as "bait" to entice the enemy aircraft into pursuit. The enemy aircraft, reacting to the bait, was drawn into the blue aircraft's weapon attack zone. This coordinated maneuver enabled the blue team to secure a positional advantage, effectively confining the red aircraft within their attack zone. This tactic not only demonstrated the effectiveness of team coordination but also shown the blue team's ability to manipulate the enemy's actions through strategic deception.

\subsection{Ablation Study}
\begin{figure}[htbp]
    \centering
    \begin{subfigure}[t]{0.45\linewidth}
        \centering
        \includegraphics[width=\linewidth]{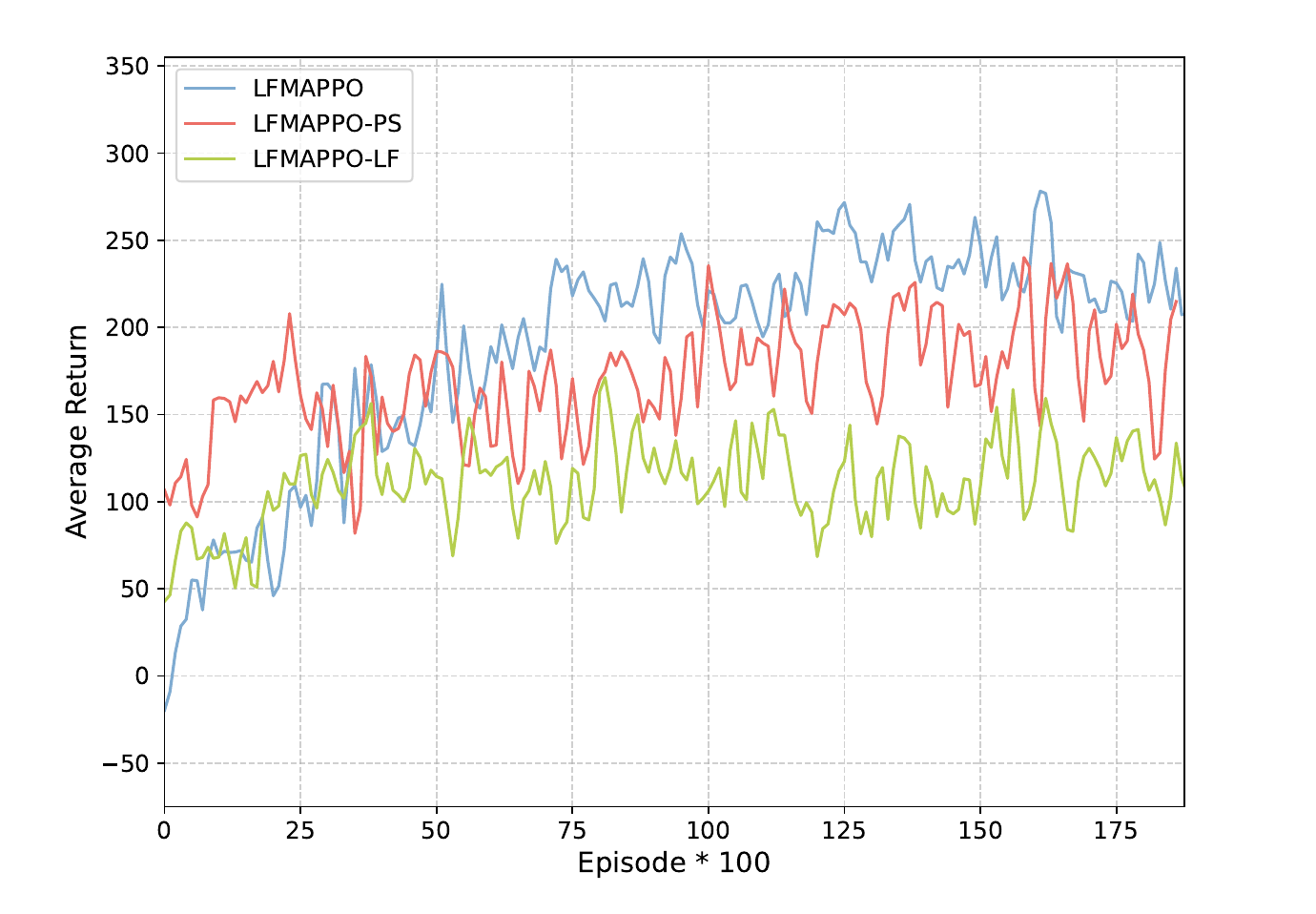}
        \caption{
        Average return for ablation on different modules}
        \label{fig:abl:module}
    \end{subfigure}
    \hfill
    \begin{subfigure}[t]{0.45\linewidth}
        \centering
        \includegraphics[width=\linewidth]{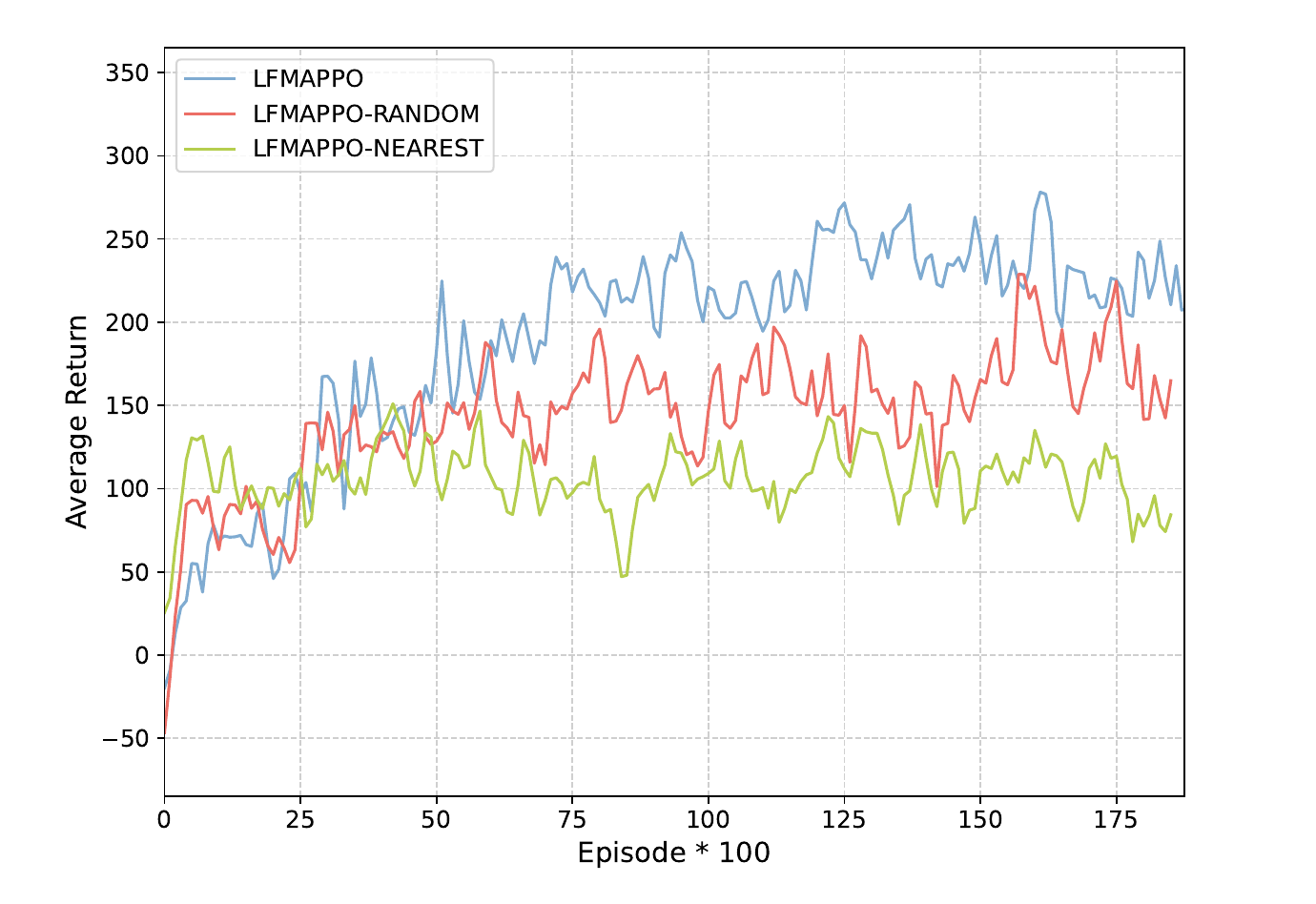}
        \caption{Average return for different target selectors}
        \label{fig:abl:target}
    \end{subfigure}
    \caption{The performance of different modules and target selectors on the average return of combat task.}
    \label{fig:}
\end{figure}

\textbf{Modules effect.}
As shown in Figure~\ref{fig:abl:module}, LFMAPPO (blue curve) demonstrates the best performance throughout the training process, consistently achieving higher reward values overall. Except for differences in modules, all other parameters remain the same.
The leader-follower strategy and hierarchical architecture play a pivotal role in enhancing the algorithm's overall performance. These modules provide a structured approach for agents to collaborate more effectively, enabling dynamic responses to changing combat conditions.
LFMAPPO-PS (red curve) removes the top layer of the hierarchical structure but retains the leader-follower strategy value function. The cooperative strategies enabled by the leader-follower strategy module yield superior results compared to the non-cooperative approach. Despite the absence of the policy selection module, the red line continues to maintain relatively high reward values. This resilience can be attributed to the foundational substructure of the framework, which is capable of generating basic movement actions and ensuring a baseline level of performance.
The average return of LFMAPPO-LF (green curve), which removes the leader-follower strategy module, is lower than the others, indicating that the cooperative policy with the leader-follower strategy achieves better results than the non-cooperative approach in combat scenarios.

\textbf{Replace the target selector.}
To assess the effectiveness of the proposed target selector, we compared it with both the random and nearest-distance target selection methods. The reward values for each method are shown in Figure~\ref{fig:abl:target}. The proposed target selector consistently achieved the highest average return. Notably, the random target selector outperformed the nearest-distance approach, as distance-based target selection can place the aircraft in vulnerable situations, leading to lower average return. Consequently, the proposed target selector evaluates the UAV's situational context, balancing both its posture and distance to the target in the decision-making process.

\section{Conclusion}
In this paper, we address the high-dimensional and collaborative challenges inherent in multi-UAV air combat by proposing a hierarchical framework based on the Leader-Follower Multi-Agent Proximal Policy Optimization strategy. We optimize the agents' value functions by assigning distinct roles to UAVs within the top-level policy selector. A target selector is introduced to evaluate threat levels based on flight status and posture. Simulation experiments validate the effectiveness of our method. Future work will focus on further refining coordination mechanisms, smoothing trajectories, and addressing the issue of agent laziness caused by relative positioning control.

\section*{Acknowledgments}
This work was supported by the Beijing Natural Science Foundation under Grant No.L247008.

\bibliographystyle{cas-model2-names}

\bibliography{refs}

\end{document}